\documentclass[
showpacs,
floatfix,
aps,
prb,
amsmath,
twocolumn,
tightenlines
groupaddress,
]{revtex4}
\usepackage{amsmath,amssymb}
\usepackage{amscd, latexsym}
\usepackage{mathrsfs}
\usepackage{graphicx} 
\usepackage{epstopdf} 
\usepackage{cancel} 
\usepackage{amsfonts}
\usepackage{exscale}
\usepackage{dcolumn}
\usepackage{bm}
\usepackage{color}
\usepackage{comment}
\usepackage{calligra}

\usepackage{calligra}
\DeclareMathAlphabet{\mathcalligra}{T1}{calligra}{m}{n}
\DeclareFontShape{T1}{calligra}{m}{n}{<->s*[2.2]callig15}{}
 
\newcommand{\be}{\begin{equation}}
\newcommand{\ee}{\end{equation}}
\newcommand{\bea}{\begin{eqnarray}}
\newcommand{\eea}{\end{eqnarray}}

\DeclareMathOperator{\Img}{Im}

\begin{document}

\title{Relaxation in quantum dots due to evanescent-wave Johnson noise}
\author{Amrit Poudel, Luke S. Langsjoen, Maxim G. Vavilov, and Robert Joynt}
\affiliation{Department of Physics, University of Wisconsin-Madison, Wisconsin 53706, USA }

\date{\today}

\begin{abstract}
We present our study of decoherence in charge (spin) qubits due to evanescent-wave Johnson noise (EWJN) in a laterally coupled double quantum dot (single quantum dot). The high density of evanescent modes in the vicinity of metallic gates causes energy relaxation and a loss of phase coherence of electrons trapped in quantum dots. We derive expressions for the resultant energy relaxation rates of charge and spin qubits in a variety of dot geometries, and EWJN is shown to be a dominant source of decoherence for spin qubits held at low magnetic fields. Previous studies in this field approximated the charge or spin qubit as a point dipole. Ignoring the finite size of the quantum dot in this way leads to a spurious divergence in the relaxation rate as the qubit approaches the metal. Our approach goes beyond the dipole approximation and remedies this unphysical divergence by taking into account the finite size of the quantum dot. Additionally, we derive an enhancement of EWJN that occurs outside a thin metallic film, relative to the field surrounding a conducting half-space.

\end{abstract}


\maketitle

\section{Introduction}
Semiconducting quantum dots are promising candidates for scalable quantum information processing~\cite{Loss98}.  Several experiments performed on laterally coupled double quantum dots (DQDs) have demonstrated precise and rapid control of the coupling between electronic charge states and coherent manipulation of trapped electrons,~\cite{Hayashi03, Petta04, Gorman05}  leading to realization of a DQD as a qubit.  Quantum dots are realized in a variety of experimental setups, including a Si and GaAs two-dimensional electron gas,~\cite{Hayashi03, Petta04, Gorman05} semiconductor nanowires~\cite{Hu07} and carbon nanotubes.~\cite{Mason04} In almost all of these implementations, confinement and manipulation of an electron in a quantum dot is achieved by applying an electrostatic potential through metallic gates. While the metallic gates are crucial for qubit control, they can also act as a source of decoherence during qubit operations.

Several decoherence mechanisms, such as hyperfine coupling of the trapped electron spin to host lattice nuclear spins in spin-based qubits~\cite{Taylor07}  and electron coupling to phonon modes~\cite{Stavrou05, Thorwart05, Vorojtsov05} in charge-based qubits, have been previously studied in an effort to identify the major source of decoherence in semiconductor qubits. A more recent study investigated decoherence due to voltage fluctuations in the metallic gates using the lumped circuit model of a DQD charge qubit.~\cite{Valente10} In almost all of these studies,~\cite{Stavrou05, Thorwart05, Vorojtsov05, Valente10} the estimated energy relaxation rate is at least an order of magnitude smaller than the rate observed experimentally,~\cite{Petta04, Hayashi03} suggesting that a different decoherence mechanism is dominant in current experimental setups for charge qubits.

Here we present our study of decoherence in a quantum dot due to electromagnetic field fluctuations near the metallic gates.  We focus primarily on noise from the high density of evanescent modes in the vicinity of metallic gates. This evanescent-wave Johnson noise (EWJN) has been identified as an important source of decoherence in atomic~\cite{Henkel99, Harber03} and quantum dot based qubits.~\cite{Langsjoen12} Our previous work \cite{Langsjoen12} as well as other theoretical estimates \cite{Henkel99} of the effect of Johnson noise in atomic and quantum dot based qubits use the dipole approximation, which is a valid approximation if the distance from the metallic gate to the qubit is much larger than the size of the qubit. However, it may be necessary to go beyond the dipole approximation in the case of EWJN in a quantum dot.

In this work, we present our study of the energy relaxation of a single electron charge qubit in a DQD system and a single electron spin qubit in a single quantum dot. We assume that the primary source of field fluctuations are the metallic top gates of the quantum dot architecture. Back gates are typically a distance on the order of a micron from the qubits, which is too far to experience significant EWJN enhancement. We consider the detailed spatial variation of the electromagnetic field fluctuations and present results beyond the dipole approximation which take into account the finite size of the quantum dot. We show that this extension of the dipole approximation removes the unphysical divergence in the relaxation rate at the metallic surface. This paper is organized as follows: In Section \ref{sec:charge} we present our formalism for calculating the relaxation rate of a charge qubit. Results are presented for a DQD geometry.  Section \ref{sec:spin} presents the formalism and results for the relaxation rate of a spin QD. In Section \ref{sec:finitethickness} we derive an enhancement of the noise spectrum that results as the thickness of the metallic gate is decreased. Finally, Section \ref{sec:conclusions} summarizes our results.  Our results indicate that EWJN is the dominant cause of energy relaxation in some spin qubit experiments, particularly those performed in a small external magnetic field, and is comparable in effect with previously studied noise sources in charge qubits.

\section{Charge qubit}
\label{sec:charge}
We consider a charge qubit realized in a gated lateral DQD in an AlGaAs/GaAs heterostructure where electron confinement along the $z$ direction is much smaller than in the $x$ or $y$ directions, so that we can safely decouple the dynamics along $x$ and $y$ directions from the $z$ direction~\footnote{In the 1D case of a DQD realized in carbon nanotube~\cite{Mason04} or semiconducting nanowire,~\cite{Hu07} we assume the electron dynamics along the $x$ direction are decoupled from that of the $y$ and $z$ directions.}. The total Hamiltonian of the charge qubit and its interaction with the electromagnetic environment is given by
\begin{equation}
\label{totalH}
H = H_q + H_{int} 
\end{equation}
where $H_q$ is the Hamiltonian of the charge qubit in a DQD, which we model in the basis of the localized charge states $\{ |L\rangle, |R\rangle\}$ as $H_q=\varepsilon/2(|L\rangle\langle L|-|R\rangle\langle R|)+\Delta/2(|L\rangle\langle R|+|R\rangle\langle L|)$. $\varepsilon$ is the bias energy between the two dots, and $\Delta$ is the tunneling amplitude. In the energy eigenbasis this Hamiltonian reduces to
\be
H_q =\frac{\hbar\omega}{2}\sigma_z
\ee 
where $\sigma_z$ is the Pauli matrix, and $\hbar\omega=\sqrt{\varepsilon^2+ \Delta^2}$.  For all our calculations except those in Fig.~\ref{figBias1D}, we will set $\varepsilon=0$.  The interaction Hamiltonian $H_{int}$ may be expressed in this same basis as

\begin{align}
\label{H-SB2} 
H_{int} &= -  \int d\vec{r}\,\Big[\hat{\sigma}_x \, \vec{M}_r(\vec{r}) +\hat{ \sigma}_z\, \vec{M}_\phi(\vec{r}) \Big]\cdot \vec{A}(\vec{r},t)~,
\end{align}
where $\vec{A} (\vec{r}, t)$ is the vector potential of the fluctuating field. $\vec{M}_r$ and  $\vec{M}_{\phi}$ are associated with energy relaxation and pure dephasing in the charge qubit, respectively and are defined as
\begin{align}
\vec{M}_r(\vec{r}) \equiv &\frac{e}{m c} \psi^*_{+}(\vec{r}) \,\vec{p} \, \psi_{-} (\vec{r}) - \frac{ie\hbar}{2m c} \psi^*_{+}(\vec{r}) \, \psi_{-} (\vec{r})\, \nabla \quad , \nonumber \\
\vec{M}_\phi(\vec{r}) \equiv & \frac{e}{2m c} \Big[\psi^*_{+}(\vec{r}) \,\vec{p} \, \psi_{+} (\vec{r}) - \psi^*_{-}(\vec{r}) \,\vec{p} \, \psi_{-} (\vec{r}) \Big] \nonumber \\
-&\frac{ie\hbar}{4 m c} \Big[\psi^*_{+}(\vec{r})\,\psi_{+} (\vec{r})\,\nabla - \psi^*_{-}(\vec{r})\,\psi_{-} (\vec{r})\,\nabla \Big] 
\end{align}
Here $m$ is the effective mass and $\vec{p}$ is the momentum operator of the trapped electron. Because we are operating within the weak field limit, the term proportional to $\vec{A}^2$ has been dropped from the interaction Hamiltonian. We choose the gauge where the scalar potential $\phi =  0$ so that $\vec{E} =- \partial_t \vec{A}$. The expression for $H_{int}$ derives from an interaction in terms of operator quantities of the form

\be
\label{Hint}
 H_{int} = -\frac{e}{2 m c}\Big( \vec{A} (\vec{r}, t)\cdot \vec{p} + \vec{p} \cdot \vec{A}(\vec{r}, t)\Big) \,.
\ee

This symmetrized version of the vector potential is not strictly necessary in our case since our qubit resides in the vacuum where $\nabla\cdot\vec{A}=0$, but we included it to keep our results more generally applicable. The zero temperature relaxation rate $\Gamma_{1,c}=1/T_{1,c}$ can be calculated using the following expression, which follows directly from Fermi's golden rule:
\begin{align}
\label{RelaxRate}
\Gamma_{1,c} &=\frac{1}{\hbar^2} \sum_{ij} \int d^3\vec{r} \int d^3\vec{r}\tiny~' \, M^{\ast i}_r (\vec{r}) \, M^j_r (\vec{r}\tiny~') \chi_{ij} (\vec{r}, \vec{r}\tiny~', \omega).
\end{align}
At finite temperature, the emission (transition from excited to ground state) and absorption (transition from ground to excited state) rates are given by 
\begin{align}
\label{gammaea}
\Gamma^e_{1,c}&=(1+N(\omega,T))\Gamma_{1,c}~\,,~\,\Gamma^a_{1,c}&=N(\omega,T)\Gamma_{1,c}
\end{align}
The Planck function $N(\omega, T) = 1/[\exp(\hbar \omega/k_BT) - 1]$ gives the average occupation number of environment modes with frequency $\omega$ at temperature $T$. The spectral density of the vector potential $\chi_{ij}$ is related to the retarded photon Green's function $D_{ij}$ by~\cite{LifshitzBook}
\begin{align}
\chi_{ij} \left(\vec{r},\vec{r}\tiny~',\omega\right) &\equiv \int d\tau \exp^{-i\omega\tau}\langle \left[A_{i}\left(\vec{r},t\right),  A_{j}\left(  \vec{r}\tiny~' ,t+\tau\right)\right] \rangle \nonumber\\&= -\frac{1}{\epsilon_0}\operatorname{Im}D_{ij}\left(  \vec{r},\vec{r}\tiny~',\omega\right)\,,\nonumber
 \end{align}
where $i,j$ are Cartesian indices that run over $x, y, z$ and the square brackets denote the commutator. $D_{ij}$ is obtained by solving
\begin{align}
\label{difeqg}
&\left[  -\delta_{ij}\left(  \nabla^{2}+\frac{\omega^{2}\epsilon\left(
\vec{r},\omega\right)  }{c^{2}}\right)  +\partial_{i}\partial_{j}\right]
D_{ik}\left(  \vec{r},\vec{r}\tiny~', \omega\right)\nonumber \\&  = -4\pi\hbar~\delta^{3}\left(
\vec{r}-\vec{r}\tiny~' \right)  \delta_{jk}.
\end{align}
Here the relative permittivity $\epsilon (\vec{r}, \omega)$ characterizes the geometry of a particular problem.  In this section, we shall limit ourselves to the case where the metallic top gate of the lateral DQD is approximated by the half-space, $z<0$. Then we can derive an analytical expression for $D_{ij}$~\cite{LifshitzBook, Agarwal75}
\begin{align}
\label{Dij}
D_{ij} (\vec{r}, \vec{r}\tiny~', \omega) = & \frac{1}{4\pi^2}\int e^{i \vec{k} \cdot\vec{r}_\parallel} \tilde{D}_{ij} (\vec{k}, z, z', \omega) \, d\vec{k}\,,  \\
\label{Dxx}
\tilde{D}_{xx} (\vec{k}, z, z', \omega)  = &\frac{2\pi i \hbar}{q} e^{iq(z+z')}\nonumber \\
\times&\Bigg[ r_s(k, \omega)\cos^2\theta - \frac{q^2c^2}{\omega^2}r_p(k, \omega)\sin^2 \theta\Bigg], 
\end{align}
where $r_s$ and $r_p$ are Fresnel's reflection coefficients given by
\begin{align}
r_p (k, \omega) = \frac{\epsilon q-q_1}{\epsilon q+q_1} \,, \quad \nonumber 
r_s(k, \omega) = \frac{q-q_1}{q+q_1} \,. \nonumber 
\end{align}
Here $\vec{k} \equiv (k_x, k_y)$, $\vec{r}_{\parallel} \equiv (x-x', y-y')$,  $\theta$ is the angle between $\vec{k}$ and the $x$-axis, and $q=\sqrt{\omega^2/c^2-k^2}$ and $q_1=\sqrt{\epsilon\omega^2/c^2-k^2}$ are the $z$-components of the photon wavevector in the vacuum and the metal, respectively. All other components of $\tilde{D}_{ij}$ can be derived from $\tilde{D}_{xx}$.~\cite{LifshitzBook} In this work we consider $\epsilon \approx i\sigma/\omega\epsilon_0$, where $\sigma = 6\times10^7 $ S/m is the conductivity of the copper gate.

We pause briefly to mention that typical models of the interaction of a DQD with the electromagnetic field use the dipole interaction Hamiltonian
\be
\label{HintD}
H_{int}  = - \vec{E}(\vec{r})\cdot\vec{d},
\ee
which will result in a relaxation rate of
\be
\Gamma_{1,c}=\frac{d^2\omega}{4\hbar z^3\sigma}
\ee
in the quasistatic approximation, where $\vec{d}$ is the dipole moment of the qubit and $\vec{E} (\vec{r})$ is the strength of the fluctuating electric field evaluated at the location of the qubit. This expression approximates that the electric field is uniform over the spatial extent of the qubit, which is equivalent to treating the qubit as a point dipole.  As such, the qubit is able to couple to arbitrarily small wavelengths of the electromagnetic spectrum, and the relaxation rate is seen to diverge at shorter distances as $\sim 1/z^3$ if the conductor is modeled with a local dielectric function~\cite{Henkel99, Langsjoen12}. Using the complete electromagnetic interaction Hamiltonian (\ref{Hint}) accounts for fluctuations of the field over the spatial extent of the qubit. If the wavelength of a particular Fourier component of the field fluctuations is smaller than the length of the qubit in that direction, its influence on the electron will average out and it will not contribute to qubit relaxation.  The exact and dipole approximation forms of the interaction Hamiltonian, Eqs. (\ref{Hint}) and (\ref{HintD}), converge when the distance from the gate becomes larger than the spatial extent of the qubit.

\begin{figure} [t] 
\includegraphics[width = 1.0 \columnwidth] {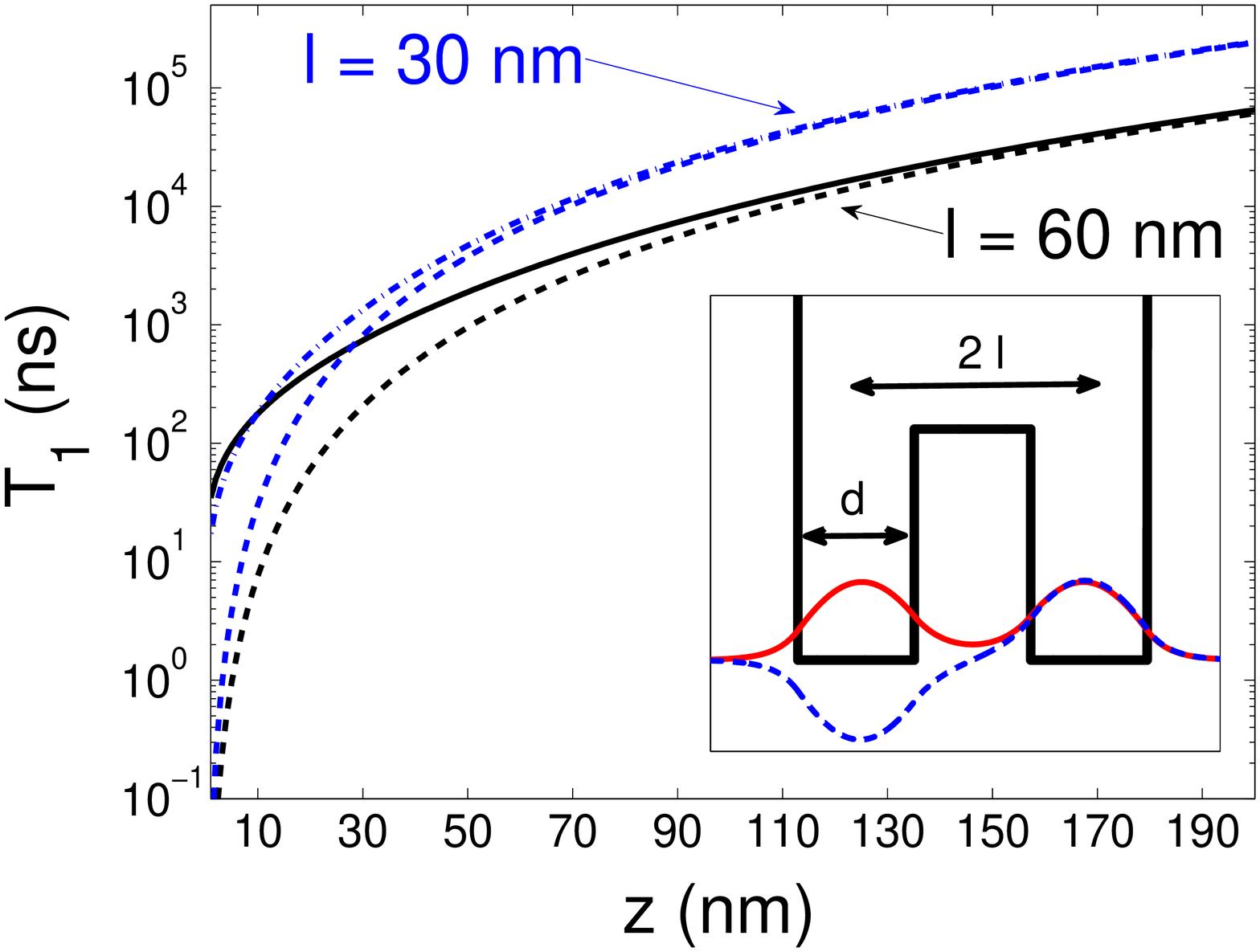}
\caption {Energy relaxation time $T_1$ vs$.$ the distance from the metallic gate to the DQDs $z$ for dot geometry of $d = 30$ nm and $l = 30$ nm (dashed and dash-dotted blue lines) and 60 nm (solid and dashed black lines) at $0$ K, $\omega/ 2 \pi  =1\,GHz$, and $\varepsilon=0$.  Solid and dash-dotted lines are $T_1$ times for the exact form of the interaction Hamiltonian, whereas dashed lines are for the dipole form of the interaction. The inset in the figure displays confining potential of a typical DQD in a one dimensional nanowire and the corresponding symmetric ground state and antisymmetric first excited state.}
\label{fig1D} 
\end{figure}

\begin{figure} 
\includegraphics[width = 1.0 \columnwidth] {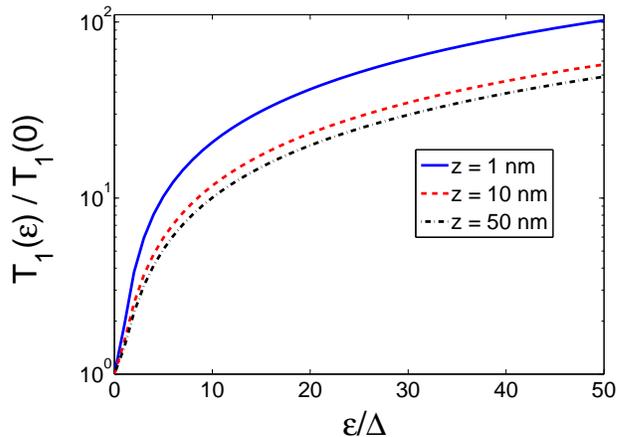}
\caption {Ratio of  the $T_1$ time for finite bias $\varepsilon$ to the $T_1$ time at zero bias vs$.$ $\varepsilon/\Delta$ for three values of distances $z$ from the metallic gate: 1 nm (solid blue line), 10 nm (dashed red line) and 50 nm (dash-dotted black line), for a dot geometry of $d = 30$ nm and $l = 30$ nm at 0 K temperature.}
\label{figBias1D} 
\end{figure}

We present calculations of the relaxation time for charge qubits that highlight the differences between these two forms of the interaction. First we consider DQDs in a one dimensional nanowire, which are realizable in semiconducting nanowires~\cite{Mason04} or carbon nanotubes~\cite{Hu07}. In such a geometry, the wave functions of trapped electrons in quantum dots have appreciable spatial extent in only one direction. We model the confining potential of the DQD as a symmetric double square well potential and compute the lowest two eigenenergies and wavefunctions. We then compute the relaxation rate between these two lowest states which are separated by a fixed transition frequency $\omega/2\pi=1GHz$. A plot of the wave functions and the shape of the potential is shown in the inset of Fig.~\ref{fig1D}.

 We plot the energy relaxation time $T_1$ vs$.$ the distance $z$ from the metallic gate to the DQD in Fig.~\ref{fig1D}. In this plot, we choose the size of the dot in the $x$-direction $d=30$ nm and half the separation between the dots $l=30$ nm (dash-dotted and dashed blue lines) and 60 nm (solid and dashed black lines). The curves that are shown in solid and dash-dotted lines are relaxation times for the exact form of the interaction Hamiltonian, whereas those shown in dashed lines are obtained using the dipole form of the interaction.  The curves show significant deviation of the exact relaxation rate from the dipole relaxation rate at shorter distances and convergence of the two results at longer distances.

\begin{figure} 
\includegraphics[width = 1.0\columnwidth] {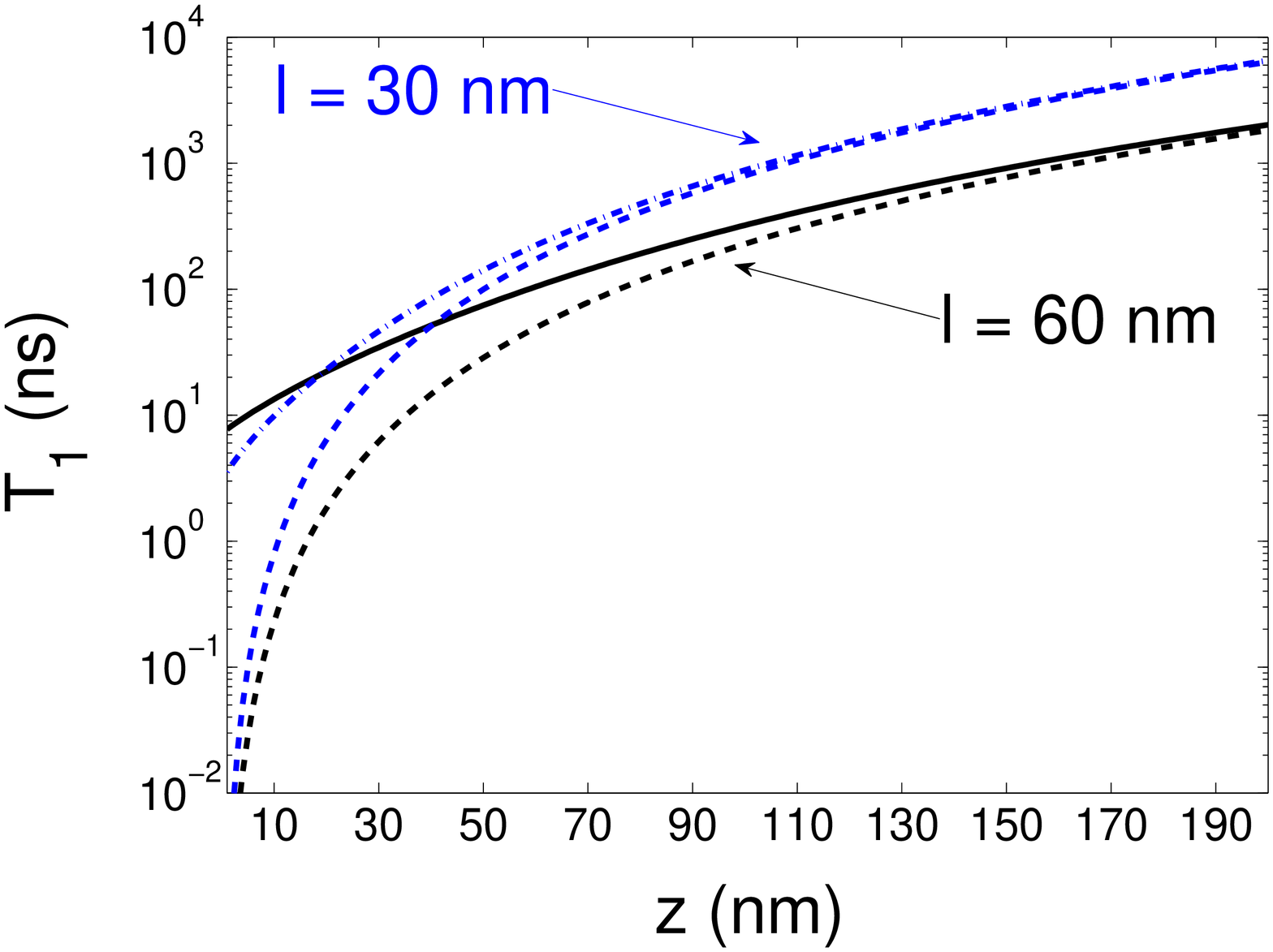}
\caption {Energy relaxation time $T_1$ vs$.$ the distance from the metallic gate to the DQDs $z$ for dot geometry of $d =  f = 30$ nm and $l = 30$ nm (dashed and dash-dotted blue lines) and 60 nm (solid and dashed black lines) at $0$ K, $\omega/2\pi = 1\,GHz$, and $\varepsilon=0$.  Solid and dash-dotted lines are $T_1$ times for the exact form of the interaction Hamiltonian, whereas dashed lines are for the dipole form of the interaction.}
\label{fig2D}
\end{figure}

In Fig.~\ref{figBias1D}, we present the ratio of $T_1$ for a charge DQD qubit at bias $\epsilon$ to the $T_1$ obtained at $\epsilon=0$ versus the ratio $\epsilon/\Delta$.  An increase in bias increases the level splitting and decreases the dipole moment of the DQD. Since the relaxation time $T_1 \sim 1/\omega d^2$, where $d=2l \sin\left( \arctan\left( \Delta/\varepsilon\right)\right)$ is the dipole moment of the quantum dot, $T_1$ increases for larger bias.

Next, we present results from the relaxation rate calculation for a DQD in a two-dimensional quantum well. In this treatment we label the $z$-axis as the vertical confinement direction and do not consider excitations along the $z$-direction. We model the confining potential by a symmetric double rectangular well in 2D and numerically compute the lowest two eigenenergies and wavefunctions.  We then compute the electron relaxation rate between these two lowest states. The results are qualitatively similar to the one-dimensional case and are shown in Fig.~\ref{fig2D}, where we plot the energy relaxation time $T_1$ vs$.$ the distance $z$ from the metallic gate to the DQD. In this plot, we choose the size of the dot in the $x$-direction $d = 30$ nm, the size in the $y$-direction $f = 30$ nm and half the separation between the dots $l$ to be 30 nm (dashed and dash-dotted blue lines) and 60 nm (solid and dashed black lines). We find that for $l = 30$ nm and $z = 90$ nm, $T_1$ is  $4~\mu s$ while for $l = 60$ nm, $T_1$ is $1.6 ~\mu s$. These relaxation times are somewhat longer than the experimentally reported value  of $T_1 = 20$ ns  in DQD-based charge qubits.~\cite{Petta04} We note that the relaxation rate for a two-dimensional DQD is shorter than for a one-dimensional DQD of comparable geometry by about a factor of 5.  A two-dimensional DQD is able to couple to obliquely oriented wavevectors in addition to those which point in the direction of separation between the dots, and this can be reasonably expected to enhance relaxation by a geometric factor of order unity. 

\section{Spin qubit}
\label{sec:spin}
We now focus on the calculation of the relaxation rate for a single electron in a spin qubit realized in a single quantum dot. Here the system $H_s$ and the interaction $H_{int}$ Hamiltonians are given by  
\begin{align}
H_s &= -g \mu_B\,\vec{\sigma} \cdot \vec{B}_0/2 \\
H_{int}&=-g \mu_B\,\vec{\sigma} \cdot \vec{B}(\vec{r}, t)
\end{align}
where $\vec{\sigma}$ is the vector of Pauli matrices, $g$ is the gyrometric  factor of the trapped electron in a quantum dot, $\mu_B$ is the Bohr magneton, $\vec{B}_0$ is the externally applied magnetic field and $\vec{B}(\vec{r}, t)$ is the fluctuating EWJN field. The rate of spin flip from excited $|\uparrow\rangle$ to ground $|\downarrow\rangle$ at $T=0K$ can be obtained from Fermi's golden rule
\begin{align}
\label{SpinRelaxRate}
\Gamma_{1, s}& = \frac{1}{\hbar^2} \int d^3\vec{r} \int d^3\vec{r}\tiny~'M_{r,s} (\vec{r}) \, M_{r, s} (\vec{r}\tiny~') \nonumber\\&\times\epsilon_{ijk}\epsilon_{ij'k'} \chi_{kk'}^B (\vec{r}, \vec{r}\tiny~', \omega)n_jn_{j'}\nonumber\\
\end{align}
where repeated indices are summed over, and $n_j$ are the components of a unit vector $\hat{n}$ in the direction of $\vec{B}_0$. The effect of finite temperature on the transition rates is the same as for charge qubits, as shown in Eq. (\ref{gammaea}). The magnetic spectral density $\chi_{ij}^B$ and the matrix element $M_{r,s}(\vec{r})$ are
\begin{align}
\label{chiB}
\chi^B_{ij}(\vec{r}, \vec{r}\tiny~',  \omega) &= \frac{\varepsilon_{ikm} \varepsilon_{jnp}}{\epsilon_0 c^2} \partial_k \partial_n \Img\, D_{mp}(\vec{r}, \vec{r}\tiny~',\omega)  \\
\label{mrs}
M_{r, s} (\vec{r}) &\equiv g \mu_B |\psi_0(\vec{r})|^2 \,. 
\end{align}
Here the spin qubit frequency $\omega=g\mu_B B_0/\hbar$ and $\psi_0(\vec{r})$ is the spatial part of the the ground state wave function of the spin qubit. Equation (\ref{SpinRelaxRate}) is a generalization beyond the dipole approximation of the simpler expression\cite{Langsjoen12}:
\be
\label{spinratedipole}
\Gamma_{1, s}= \frac{g^3\mu_B^3\sigma B_0}{8\hbar^2\epsilon_0c^4z},
\ee
which has been obtained by using the quasistatic limit for the Green's function (\ref{chiB}), and assuming it is constant over the spatial extent of the qubit. Equation (\ref{spinratedipole}) also assumes the external magnetic field $\vec{B}_0$ points in the $z$-direction.

\begin{figure} 
\includegraphics[width = 1.0\columnwidth] {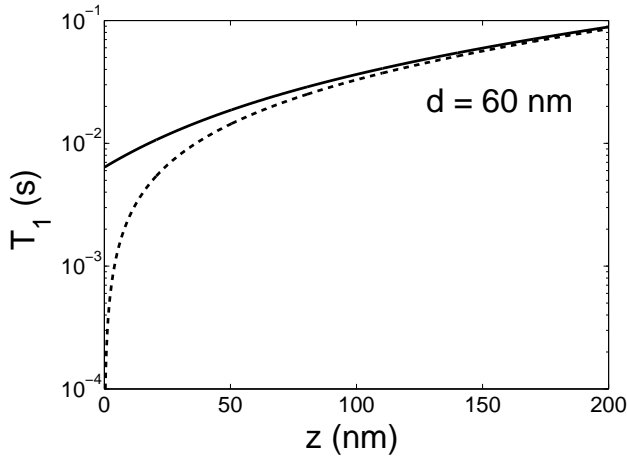}
\caption {Energy relaxation time $T_1$ vs$.$ the distance from the metallic gate to quantum dot $z$ for spin qubit single dot geometry of $d =  60$ nm at 0 K temperature and $g\mu_B B_0/2\pi\hbar = 50\,GHz$.  The solid line represent $T_1$ time for the exact form of the interaction Hamiltonian, whereas the dashed line is for the dipole form of the interaction.}
\label{figspinQD}
\end{figure}

A plot of energy relaxation time $T_1$ $vs.$ the distance from the metallic gate $z$ for a spin qubit is displayed in Fig.~\ref{figspinQD}.  Here we consider a single quantum dot of diameter $d = 60$ nm and approximate the ground state spatial wave function of the spin qubit by the ground state wave function of a harmonic potential. We assume the Zeeman splitting between spin states is $50\,GHz$, typical of experiments in spin qubits.~\cite{Elzerman04} The solid line is the $T_1$ time obtained using the non-local magnetic spectral density while the dotted line is obtained for a local spectral density, which diverges as $\sim 1/z$ as one approaches the metallic gate. The reason for saturation of the $T_1$ time at smaller distances is similar to the case for charge qubits. There is a slight distinction in that the spin case involves a spatially extended dipole interaction, as opposed to the charge case which involves qenuine quadrupole and higher multipole contributions.  This distinction is largely technical, however, and a saturation of $T_1$ as $z\rightarrow0$ is observed in both cases. We find that the $T_1$ time for a spin qubit in a GaAs quantum dot with an external magnetic field of $10\,T$ and $z = 90$ nm is 150 ms which is larger than the experimentally reported value \cite{Elzerman04} of 0.55 ms, and generally EWJN does not seem to be the dominant source of decoherence for semiconductor devices in large magnetic fields.  GaAs has a strong spin-orbit interaction (SOI), which mixes the Zeeman-split spin states with orbitally excited states.  Spin relaxation can then occur via coupling of the qubit to piezoelectric phonon noise in the 2DEG layer. The relaxation rate from this mechanism scales as $B^5$ and is the dominant pathway for spin relaxation at large external magnetic fields $B>1T$\cite{Amasha, Khaetskii}.  Additionally, Marquardt and Abalmassov \cite{Marquardt} calculate relaxation of spin qubits from electric EWJN via the SOI.  Again, mixing of the charge and spin states via the SOI allows spin relaxation to be induced from electric field fluctuations. They estimate the power spectrum of the Johnson noise using a lumped circuit model and found a $B^3$ dependence of the relaxation rate.  Our treatment involves a direct coupling of the fluctuating magnetic field from the top gates with the spin states, and our rate scales linearly with the magnetic field.  We therefore expect our relaxation pathway to dominate at low magnetic fields, and indeed while we predict a much slower relaxation rate than measured by Amasha et al \cite{Amasha} for $B\sim 7T$, at $B=1T$ our results predict $T_1\sim1.5$ s which is comparable to their measured value of $T_1=1$ s.  Additionally, in Si quantum dots with a $2\,T$ external magnetic field and $z = 50$ nm, we predict a $T_1$ time of 6 ms which is smaller than the experimentally reported value of 40 ms.~\cite{Xiao10} However, it must be kept in mind that we have so far considered the simpler top gate geometry of a conducting half-space rather than the thin layer of finger gates used in these experiments.  In the next section we address modifications to our calculations that we expect from more realistic gate geometry.

\section{Thin metallic gates}
\label{sec:finitethickness}
A conducting half-space is an analytically convenient gate geometry, but a poor approximation to the thin top gates commonly used in semiconductor devices.  In this section we present an exact treatment of the behavior of EWJN in the vicinity of a metallic film of finite thickness.   Changing the half-space to a thin film affects EWJN by modifying the reflection coefficients $r_s$ and $r_p$. The power spectrum of the resultant EWJN is obtained by substituting these modified reflection coefficients into the photon Green's function (\ref{Dxx}), and the relaxation time of, e.g. a charge qubit, is obtained by plugging Eq. (\ref{Dxx}) into Eqs. (\ref{RelaxRate}) and (\ref{Dij}). The modified reflection coefficients for a film of thickness $a$ take the form
\begin{align}
\label{rpt}
r_p (k, \omega,a) &= \frac{\epsilon^2 q^2-q_1^2 }{q_1^2+\epsilon^2q^2 +2iqq_1\epsilon \cot(q_1a)} \\
\label{rst}
r_s (k, \omega,a) &= \frac{q^2-q_1^2}{q^2+q_1^2+2iqq_1\cot(q_1a)}.
\end{align} 
They differ significantly from the half-space result only when the thickness $a$ is of the order or smaller than the skin depth $\delta$, and they reduce to the half-space result for $a\gg\delta$. A derivation of Eqs. (\ref{rpt}) and (\ref{rst}) is given in the Appendix. Equations (\ref{rpt}) and (\ref{rst}) are exact, but for a good conductor they can be cast into a simpler approximate form
\begin{align}
\label{rptqs}
r_p(k,\omega,a) &\approx \left(1+\frac{2q_1}{\epsilon k}\cot\left(q_1a\right)\right)^{-1}\\
\label{rstqs}
r_s(k,\omega,a) &\approx -\left(1-\frac{2c^2q_1k}{\epsilon\omega^2}\cot(q_1a)\right)^{-1}~.
\end{align}

These expressions have been obtained by expanding Eqs. (\ref{rpt}) and (\ref{rst}) for large imaginary $\epsilon$ and then taking the quasistatic approximation $q\rightarrow  ik$.  The first approximation is extremely accurate for copper near zero temperature and the second is accurate for all distances $z$ such that EWJN is appreciably enhanced above blackbody radiation\cite{Langsjoen12}. The remarkable feature of Eqs. (\ref{rptqs}) and (\ref{rstqs}) is that they show the strength of the fluctuating fields outside the film are actually amplified relative to the half-space result.  This can be understood by analogy to the behavior of a particle trapped in a finite one-dimensional potential well.  For a given width of the well, the wavefunction will have an exponentially decaying tail in the forbidden region.  As the confinement is increased, the particle will be squeezed and its wavefunction will leak farther into the forbidden region. It will be interesting to see if this enhancement is observable in the Casimir attraction between 2 thin conducting plates.

\begin{figure} 
\includegraphics[width = 1.0\columnwidth] {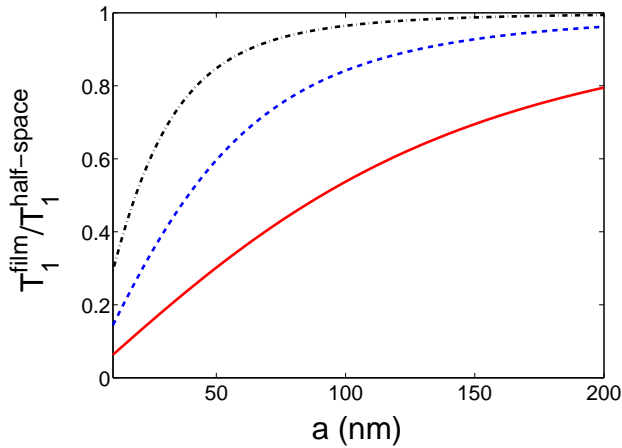}
\caption {Ratio of energy relaxation time $T_1$ from conducting film to $T_1$ time from half-space $vs.$ thickness of the film $a$ for a DQD charge qubit in one dimension with dot geometry  $d =  30$ nm  and $l = 60$ nm at 0 K temperature. We take the exact form of the interaction Hamiltonian. The distance $z$ from the film or half-space is chosen as follows: $z = 10$ nm (black dash-dotted line), $z = 50$ nm (blue dashed line) and $z= 150$ nm (solid red line). Other parameters are the same as in Fig.~\ref{fig1D}.}
\label{figSlab}
\end{figure}

Using the modified expression for the reflection coefficients, we compute the $T_1$ time of a DQD charge qubit in one dimension due to the metallic film. In Fig.~\ref{figSlab}, we plot the ratio of the $T_1$ time obtained for the film to the time computed for the metallic half-space as a function of the film thickness. We take the exact form of the interaction Hamiltonian for a variety of distances from the gate. We find that for distance $z > a$, the relaxation time due to the film can be reduced by over an order of magnitude relative to the half-space. It converges to the half-space result as $z$ becomes smaller than the thickness of the film.  

Common semiconductor qubit architectures employ thin finger-shaped top gates which are more sparse than the films considered here.  An exact treatment of EWJN from a detailed finger gate geometry would be prohibitively difficult, but we expect to a reasonable approximation that EWJN from finger gates will be reduced by a factor of the fraction of the top gate layer that is not composed of metal. Our results should then overestimate the relaxation rate by a geometric factor.  We note however that newer accumulation-mode architectures employ a second top gate above the confinement top gates\cite{Borselli}.  These accumulation gates are solid sheets and are typically around $100$ nm from the qubit, so our treatment should accurately describe their contribution to relaxation.

\section{Discussion and Conclusions}
\label{sec:conclusions}
In conclusion, we have presented a detailed study of the effect of evanescent-wave Johnson noise on energy relaxation of quantum dots beyond the dipole approximation. We have noted that previous studies of charge and spin qubits which use the dipole approximation allow contribution from infinitely large components of the photon wavevector leading to overestimation and divergence of the energy relaxation rate as $z \rightarrow 0$. We have demonstrated that it is possible to remedy this spurious divergence by taking into account the finite size of the quantum dot. While a non-local permittivity of the surface metal will remove the divergence in the field fluctuations at the boundary, we have shown that the finite size of the dot provides an alternative normalization mechanism by enforcing a finite cutoff in the magnitude of the contributing wavevector.  In addition, we have derived a novel enhancement of the EWJN field fluctuations that occurs outside a metallic film, relative to the field outside a metallic half-space.  

This manuscript has focused exclusively on relaxation, though we expect dephasing times from EWJN to be of comparable magnitude.  The power spectrum of EWJN is linear in $\omega$, and this will suppress contribution from the small frequency part of the electromagnetic spectrum, which typically enhances dephasing rates. While the temperature dependence of the relaxation rate is simply given by the Planck function, we do expect a more non-trivial temperature dependence of the dephasing rate. 

Of particular interest are experimental signatures of EWJN-induced relaxation.  Notably, at zero temperature the charge relaxation rate scales linearly with the qubit transition frequency and as the inverse cubic power of the distance between the qubit and the metallic top gates.  The zero temperature spin relaxation rate scales linearly with the external magnetic field and inversely with the distance to the gates.

Our results indicate that EWJN from the metallic top gate is not a dominant source of relaxation in charge qubits, but can be the dominant noise source for energy relaxation in spin qubits held at low external magnetic field.

\acknowledgements
We thank M.A. Eriksson for useful discussions.  This work was supported by ARO and LPS grant no. W911NF-11-1-0030 and NSF grant DMR 0955500.

 \renewcommand{\theequation}{A-\arabic{equation}}
  \setcounter{equation}{0}  
  \section*{APPENDIX}  
\subsection*{Derivation of Green's tensor for a thin film}

Here we present the calculation for the retarded photon Green's tensor outside of a thin conducting sheet of permittivity $\epsilon$. The Green's function will satisfy Eq. (\ref{difeqg}). Here $\vec{r}\tiny~'$ is simply a parameter for the purposes of solving this set of equations, and we take it to lie in the vacuum outside the conducting sheet. We will suppress the dependence of $D_{ik}(\vec{r},\vec{r}\tiny~',\omega)$ on $\vec{r}\tiny~'$ and $\omega$ to simplify the notation. The geometry of the problem is contained entirely in the permittivity function $\varepsilon\left(\vec{r}, \omega\right)$. We take the boundaries of the conducting sheet to be located at $z=-a$ and $z=0$, with vacuum outside.  Because the geometry is still translationally invariant in the $x$ and $y$ directions, we employ the same Fourier expansion (\ref{Dij}) as in Section \ref{sec:charge}. Solving Eq. (\ref{difeqg}) for a problem with planar symmetry is greatly simplified by separately considering the Fourier components of (\ref{Dij}) that are polarized in the $x$ and $y$ directions.  $\tilde{D}_{yy}(\vec{r})$ may then be reconstructed as
\begin{align}
D_{yy}(\vec{r})&=\int \frac{d^2\vec{k}}{(2\pi)^2}e^{i\vec{k}\cdot\vec{r}_\parallel}\nonumber\\&\times\left(\tilde{D}_{yy,k_x}(k,z)\cos^2\theta + \tilde{D}_{yy,k_y}(k,z)\sin^2\theta\right)
\end{align}
where $\tilde{D}_{yy,k_x}=\tilde{D}_{yy}$ when $k_y=0$, and $\tilde{D}_{yy,k_y}=\tilde{D}_{yy}$ when $k_x=0$. The boundary value problem for $\tilde{D}_{yy,k_x}(k,z)$ then becomes
\begin{align}
\tilde{D}_{yy,k_x}(k,z)= \left\{ \begin{array}{rcl}
Ae^{-iqz} & \mbox{,}
& z<-a \\ B_1e^{-iq_1z}+B_2e^{iq_1 z}  & \mbox{,} & -a\leq z<0 \\
Ce^{iqz} +\frac{2\pi\hbar}{iq}e^{iq|z-z'|}& \mbox{,} & z\geq 0
\end{array}\right.
\end{align}

Our interest lies in the behavior of the fields for $z>0$, so we need only to calculate $C$. Enforcing that $D_{yy,k_x}$ and $\partial D_{yy,k_x}/\partial z$ are continuous across the boundaries results in
\be
C= \frac{2\pi i\hbar}{q}r_s(k,\omega,a)e^{iqz'}
\ee
where
\begin{align}
\label{gammay}
r_s(k,\omega,a)&\equiv\frac{\big(q^2-q_1^2\big)\sin(q_1a)}{\big(q_1^2+q^2\big)\sin(q_1a)+ 2iqq_1\cos(q_1a)}\nonumber\\
&=2i\sin q_1a\left(e^{iq_1a}\frac{q-q_1}{q+q_1}-e^{-iq_1a}\frac{q+q_1}{q-q_1}\right)^{-1}
\end{align}
so that
\be
\tilde{D}_{yy,k_x}(k,z) = \frac{2\pi i\hbar}{q} \left(r_s(k,\omega,a)e^{iq(z+z')}  +e^{iq|z-z'|}\right)
\ee
The term proportional to $\exp(iq|z-z'|)$ is the free photon contribution to the power spectrum.  It will have an imaginary component and thus contribute to relaxation only in the radiative regime, $k\leq\omega/c$.  Within a skin depth of separation from the metal, evanescent waves are orders of magnitude larger in field strength than these free photons. They may be safely ignored in this context. A similar calculation yields the result for $\tilde{D}_{yy,k_y}$:
\be
\tilde{D}_{yy,k_y}(k,z) = - \frac{2\pi i\hbar c^2q}{\omega^2}\left( r_p(k,\omega,a)e^{iq(z+z')} -e^{iq|z-z'|}\right),
\ee
where
\begin{align}
\label{gammax}
r_p(k,\omega,a) & \equiv \frac{\big(\epsilon^2 q^2-q_1^2\big)\sin(q_1 a) }{\big(q_1^2+\epsilon^2q^2\big)\sin(q_1 a) +2iqq_1\epsilon \cos(q_1a)} \nonumber \\
 &= 2i\sin q_1a\left(e^{iq_1a}\frac{\epsilon q-q_1}{\epsilon q+q_1}-e^{-iq_1a}\frac{\epsilon q+q_1}{\epsilon q-q_1}\right)^{-1}
\end{align}
A Taylor expansion of Eqs. (\ref{gammay}) and (\ref{gammax}) for large $a$ in the evanescent range of wavevectors, i.e., a Taylor expansion in powers of $\exp(-2|q_1|a)$, gives a monotonically increasing function of film thickness, $a$.  However, a more careful treatment reveals that this is an error.  The naive expansions of (\ref{gammay}) and (\ref{gammax}) for large $a$ neglect an enhancement of the field spectrum that occurs for small $k$. In fact, EWJN is enhanced as the thickness is decreased for any good conductor. Specifically, the enhancement is preserved for a particular spatial Fourier component of the Green's function as long as $|\frac{2q_1}{\epsilon k}|<1$.  EWJN will eventually vanish as $a\rightarrow0$, but this does not occur until an unphysically small thickness is reached, on the order of $10^{-14}$ m for copper at $T=0K$ which is well below the applicability of the local permittivity model. 


\begin{thebibliography}{18}
\expandafter\ifx\csname natexlab\endcsname\relax\def\natexlab#1{#1}\fi
\expandafter\ifx\csname bibnamefont\endcsname\relax
  \def\bibnamefont#1{#1}\fi
\expandafter\ifx\csname bibfnamefont\endcsname\relax
  \def\bibfnamefont#1{#1}\fi
\expandafter\ifx\csname citenamefont\endcsname\relax
  \def\citenamefont#1{#1}\fi
\expandafter\ifx\csname url\endcsname\relax
  \def\url#1{\texttt{#1}}\fi
\expandafter\ifx\csname urlprefix\endcsname\relax\def\urlprefix{URL }\fi
\providecommand{\bibinfo}[2]{#2}
\providecommand{\eprint}[2][]{\url{#2}}

\bibitem[{\citenamefont{Loss and DiVincenzo}(1998)}]{Loss98}
\bibinfo{author}{\bibfnamefont{D.}~\bibnamefont{Loss}} \bibnamefont{and}
  \bibinfo{author}{\bibfnamefont{D.~P.} \bibnamefont{DiVincenzo}},
  \bibinfo{journal}{Phys. Rev. A} \textbf{\bibinfo{volume}{57}},
  \bibinfo{pages}{120} (\bibinfo{year}{1998}).

\bibitem[{\citenamefont{Hayashi et~al.}(2003)\citenamefont{Hayashi, Fujisawa,
  Cheong, Jeong, and Hirayama}}]{Hayashi03}
\bibinfo{author}{\bibfnamefont{T.}~\bibnamefont{Hayashi}},
  \bibinfo{author}{\bibfnamefont{T.}~\bibnamefont{Fujisawa}},
  \bibinfo{author}{\bibfnamefont{H.~D.} \bibnamefont{Cheong}},
  \bibinfo{author}{\bibfnamefont{Y.~H.} \bibnamefont{Jeong}}, \bibnamefont{and}
  \bibinfo{author}{\bibfnamefont{Y.}~\bibnamefont{Hirayama}},
  \bibinfo{journal}{Phys. Rev. Lett.} \textbf{\bibinfo{volume}{91}}, \bibinfo{pages}{226804}
  (\bibinfo{year}{2003}).

\bibitem[{\citenamefont{Petta et~al.}(2004)\citenamefont{Petta, Johnson,
  Marcus, Hanson, and Gossard}}]{Petta04}
\bibinfo{author}{\bibfnamefont{J.R.}~\bibnamefont{Petta}},
  \bibinfo{author}{\bibfnamefont{A.~C.} \bibnamefont{Johnson}},
  \bibinfo{author}{\bibfnamefont{C.~M.} \bibnamefont{Marcus}},
  \bibinfo{author}{\bibfnamefont{M.~P.} \bibnamefont{Hanson}},
  \bibnamefont{and} \bibinfo{author}{\bibfnamefont{A.~C.}
  \bibnamefont{Gossard}}, \bibinfo{journal}{Phys. Rev. Lett.}
  \textbf{\bibinfo{volume}{93}}, \bibinfo{pages}{186802} (\bibinfo{year}{2004}).

\bibitem[{\citenamefont{Gorman et~al.}(2005)\citenamefont{Gorman, Hasko, and
  Williams}}]{Gorman05}
\bibinfo{author}{\bibfnamefont{J.}~\bibnamefont{Gorman}},
  \bibinfo{author}{\bibfnamefont{D.~G.} \bibnamefont{Hasko}}, \bibnamefont{and}
  \bibinfo{author}{\bibfnamefont{D.~A.} \bibnamefont{Williams}},
  \bibinfo{journal}{Phys. Rev. Lett.} \textbf{\bibinfo{volume}{95}}, \bibinfo{pages}{090502}
  (\bibinfo{year}{2005}).

\bibitem[{\citenamefont{Hu et~al.}(2007)\citenamefont{Hu, Churchill, Reilly,
  Xiang, Lieber, and Marcus}}]{Hu07}
\bibinfo{author}{\bibfnamefont{Y.}~\bibnamefont{Hu}},
  \bibinfo{author}{\bibfnamefont{H.~O.~H.} \bibnamefont{Churchill}},
  \bibinfo{author}{\bibfnamefont{D.~J.} \bibnamefont{Reilly}},
  \bibinfo{author}{\bibfnamefont{J.}~\bibnamefont{Xiang}},
  \bibinfo{author}{\bibfnamefont{C.~M.} \bibnamefont{Lieber}},
  \bibnamefont{and} \bibinfo{author}{\bibfnamefont{C.~M.}
  \bibnamefont{Marcus}}, \bibinfo{journal}{Nature Nanotechnology}
  \textbf{\bibinfo{volume}{2}}, \bibinfo{pages}{622} (\bibinfo{year}{2007}).

\bibitem[{\citenamefont{Mason et~al.}(2004)\citenamefont{Mason, Biercuk, and
  Marcus}}]{Mason04}
\bibinfo{author}{\bibfnamefont{N.}~\bibnamefont{Mason}},
  \bibinfo{author}{\bibfnamefont{M.~J.} \bibnamefont{Biercuk}},
  \bibnamefont{and} \bibinfo{author}{\bibfnamefont{C.~M.}
  \bibnamefont{Marcus}}, \bibinfo{journal}{Science}
  \textbf{\bibinfo{volume}{303}}, \bibinfo{pages}{655} (\bibinfo{year}{2004}).

\bibitem[{\citenamefont{Taylor et~al.}(2007)\citenamefont{Taylor, Petta,
  Johnson, Yacoby, Marcus, and Lukin}}]{Taylor07}
\bibinfo{author}{\bibfnamefont{J.M.}~\bibnamefont{Taylor}},
  \bibinfo{author}{\bibfnamefont{J.R.}~\bibnamefont{Petta}},
  \bibinfo{author}{\bibfnamefont{A.~C.} \bibnamefont{Johnson}},
  \bibinfo{author}{\bibfnamefont{A.}~\bibnamefont{Yacoby}},
  \bibinfo{author}{\bibfnamefont{C.~M.} \bibnamefont{Marcus}},
  \bibnamefont{and} \bibinfo{author}{\bibfnamefont{M.~D.} \bibnamefont{Lukin}},
  \bibinfo{journal}{Phys. Rev. B} \textbf{\bibinfo{volume}{76}}, \bibinfo{pages}{035315}
  (\bibinfo{year}{2007}).

\bibitem[{\citenamefont{Stavrou and Hu}(2005)}]{Stavrou05}
\bibinfo{author}{\bibfnamefont{V.~N.} \bibnamefont{Stavrou}} \bibnamefont{and}
  \bibinfo{author}{\bibfnamefont{X.}~\bibnamefont{Hu}}, \bibinfo{journal}{Phys.
  Rev. B} \textbf{\bibinfo{volume}{72}}, \bibinfo{pages}{075362} (\bibinfo{year}{2005}).

\bibitem[{\citenamefont{Thorwart et~al.}(2005)\citenamefont{Thorwart, Eckel, and
  Mucciolo}}]{Thorwart05}
\bibinfo{author}{\bibfnamefont{M.}~\bibnamefont{Thorwart}},
  \bibinfo{author}{\bibfnamefont{J.}~\bibnamefont{Eckel}}, \bibnamefont{and}
  \bibinfo{author}{\bibfnamefont{E.~R.} \bibnamefont{Mucciolo}},
  \bibinfo{journal}{Phys. Rev. B} \textbf{\bibinfo{volume}{72}}, \bibinfo{pages}{235320}
  (\bibinfo{year}{2005}).

\bibitem[{\citenamefont{Vorojtsov et~al.}(2005)\citenamefont{Vorojtsov,
  Mucciolo, and Baranger}}]{Vorojtsov05}
\bibinfo{author}{\bibfnamefont{S.}~\bibnamefont{Vorojtsov}},
  \bibinfo{author}{\bibfnamefont{E.~R.} \bibnamefont{Mucciolo}},
  \bibnamefont{and} \bibinfo{author}{\bibfnamefont{H.~U.}
  \bibnamefont{Baranger}}, \bibinfo{journal}{Phys. Rev. B}
  \textbf{\bibinfo{volume}{71}}, \bibinfo{pages}{205322} (\bibinfo{year}{2005}).

\bibitem[{\citenamefont{Valente et~al.}(2010)\citenamefont{Valente, Mucciolo,
  and Wilhelm}}]{Valente10}
\bibinfo{author}{\bibfnamefont{D.~C.~B.} \bibnamefont{Valente}},
  \bibinfo{author}{\bibfnamefont{E.~R.} \bibnamefont{Mucciolo}},
  \bibnamefont{and} \bibinfo{author}{\bibfnamefont{F.~K.}
  \bibnamefont{Wilhelm}}, \bibinfo{journal}{Phys. Rev. B}
  \textbf{\bibinfo{volume}{82}}, \bibinfo{pages}{125302} (\bibinfo{year}{2010}).

\bibitem[{\citenamefont{Henkel et~al.}(1999)\citenamefont{Henkel, Potting, and
  Wilkens}}]{Henkel99}
\bibinfo{author}{\bibfnamefont{C.}~\bibnamefont{Henkel}},
  \bibinfo{author}{\bibfnamefont{S.}~\bibnamefont{Potting}}, \bibnamefont{and}
  \bibinfo{author}{\bibfnamefont{M.}~\bibnamefont{Wilkens}},
  \bibinfo{journal}{Applied Physics B: Lasers and Optics}
  \textbf{\bibinfo{volume}{69}}, \bibinfo{pages}{379} (\bibinfo{year}{1999}).

\bibitem[{\citenamefont{Harber et~al.}(2003)\citenamefont{Harber, McGuirk,
  Obrecht, and Cornell}}]{Harber03}
\bibinfo{author}{\bibfnamefont{D.~M.} \bibnamefont{Harber}},
  \bibinfo{author}{\bibfnamefont{J.~M.} \bibnamefont{McGuirk}},
  \bibinfo{author}{\bibfnamefont{J.~M.} \bibnamefont{Obrecht}},
  \bibnamefont{and} \bibinfo{author}{\bibfnamefont{E.~A.}
  \bibnamefont{Cornell}}, \bibinfo{journal}{Journal of Low Temperature Physics}
  \textbf{\bibinfo{volume}{133}}, \bibinfo{pages}{229} (\bibinfo{year}{2003}).

\bibitem[{\citenamefont{Langsjoen et~al.}(2012)\citenamefont{Langsjoen, Poudel,
  Vavilov, and Joynt}}]{Langsjoen12}
\bibinfo{author}{\bibfnamefont{L.~S.} \bibnamefont{Langsjoen}},
  \bibinfo{author}{\bibfnamefont{A.}~\bibnamefont{Poudel}},
  \bibinfo{author}{\bibfnamefont{M.~G.} \bibnamefont{Vavilov}},
  \bibnamefont{and} \bibinfo{author}{\bibfnamefont{R.}~\bibnamefont{Joynt}},
  \bibinfo{journal}{Phys. Rev. A} \textbf{\bibinfo{volume}{86}},
  \bibinfo{pages}{010301(R)} (\bibinfo{year}{2012}).

\bibitem[{\citenamefont{Lifshitz and Pitaevskii}(1980)}]{LifshitzBook}
\bibinfo{author}{\bibfnamefont{E.~M.} \bibnamefont{Lifshitz}} \bibnamefont{and}
  \bibinfo{author}{\bibfnamefont{L.~P.} \bibnamefont{Pitaevskii}},
  \emph{\bibinfo{title}{Statistical Physics, Part 2}}, vol.~\bibinfo{volume}{9}
  of \emph{\bibinfo{series}{Course in Theoretical Physics}}
  (\bibinfo{publisher}{Pergamon}, \bibinfo{year}{1980}).

\bibitem[{\citenamefont{Agarwal}(1975)}]{Agarwal75}
\bibinfo{author}{\bibfnamefont{G.}~\bibnamefont{Agarwal}},
  \bibinfo{journal}{Phys. Rev. A} \textbf{\bibinfo{volume}{11}},
  \bibinfo{pages}{253} (\bibinfo{year}{1975}).

\bibitem[{\citenamefont{Elzerman et~al.}(2004)\citenamefont{Elzerman, Hanson,
  Willems~van Beveren, Witkamp, Vandersypen, and Kouwenhoven}}]{Elzerman04}
\bibinfo{author}{\bibfnamefont{J.~M.} \bibnamefont{Elzerman}},
  \bibinfo{author}{\bibfnamefont{R.}~\bibnamefont{Hanson}},
  \bibinfo{author}{\bibfnamefont{L.~H.} \bibnamefont{Willems~van Beveren}},
  \bibinfo{author}{\bibfnamefont{B.}~\bibnamefont{Witkamp}},
  \bibinfo{author}{\bibfnamefont{L.~M.~K.} \bibnamefont{Vandersypen}},
  \bibnamefont{and} \bibinfo{author}{\bibfnamefont{L.~P.}
  \bibnamefont{Kouwenhoven}}, \bibinfo{journal}{Nature}
  \textbf{\bibinfo{volume}{430}}, \bibinfo{pages}{431} (\bibinfo{year}{2004}).

\bibitem[{\citenamefont{Xiao et~al.}(2010)\citenamefont{Xiao, House, and
  Jiang}}]{Xiao10}
\bibinfo{author}{\bibfnamefont{M.}~\bibnamefont{Xiao}},
  \bibinfo{author}{\bibfnamefont{M.~G.} \bibnamefont{House}}, \bibnamefont{and}
  \bibinfo{author}{\bibfnamefont{H.~W.} \bibnamefont{Jiang}},
  \bibinfo{journal}{Phys. Rev. Lett.} \textbf{\bibinfo{volume}{104}},
  \bibinfo{pages}{096801} (\bibinfo{year}{2010}).
  
\bibitem[{\citenamefont{Amasha et~al.}(2008)\citenamefont{Amasha, MacLean, Radu, Zumbuhl, Kastner, Hanson, and Gossard}}]{Amasha}
\bibinfo{author}{\bibfnamefont{S.} \bibnamefont{Amasha}},
  \bibinfo{author}{\bibfnamefont{K.}~\bibnamefont{MacLean}},
  \bibinfo{author}{\bibfnamefont{I.~P.} \bibnamefont{Radu}},
  \bibnamefont{and} \bibinfo{author}{\bibfnamefont{D.~M.}~\bibnamefont{Zumbuhl}},
  \bibnamefont{and} \bibinfo{author}{\bibfnamefont{M.~A.}~\bibnamefont{Kastner}}, 
  \bibnamefont{and} \bibinfo{author}{\bibfnamefont{M.~P.}~\bibnamefont{Hanson}}, \bibnamefont{and}
  \bibnamefont{and} \bibinfo{author}{\bibfnamefont{A.~C.}~\bibnamefont{Gossard}},
  \bibinfo{journal}{Phys. Rev. Lett.} \textbf{\bibinfo{volume}{100}},
  \bibinfo{pages}{046803} (\bibinfo{year}{2008}).

\bibitem[{\citenamefont{Khaetskii et~al.}(2001)\citenamefont{Khaetskii and Nazarov}}]{Khaetskii}
\bibinfo{author}{\bibfnamefont{A.~V.}~\bibnamefont{Khaetskii}} \bibnamefont{and}
  \bibinfo{author}{\bibfnamefont{Y.~V.} \bibnamefont{Nazarov}}, 
  \bibinfo{journal}{Phys. Rev. B.} \textbf{\bibinfo{volume}{64}},
  \bibinfo{pages}{125316} (\bibinfo{year}{2001}).
  

\bibitem[{\citenamefont{Marquardt et~al.}(2001)\citenamefont{Marquardt and Abalmassov}}]{Marquardt}
\bibinfo{author}{\bibfnamefont{F.}~\bibnamefont{Marquardt}} \bibnamefont{and}
  \bibinfo{author}{\bibfnamefont{V.~A.} \bibnamefont{Abalmassov}}, 
  \bibinfo{journal}{Phys. Rev. B.} \textbf{\bibinfo{volume}{71}},
  \bibinfo{pages}{165325} (\bibinfo{year}{2005}).


\bibitem[{\citenamefont{Borselli et~al.}(2011)\citenamefont{Borselli}}]{Borselli}
\bibinfo{author}{\bibfnamefont{M. G. Borselli, K. Eng, E. T. Croke, B. M. Maune, B. Huang, R. S. Ross, A. A. Kiselev, P. W. Deelman, I. Alvarado-Rodriguez, A. E. Schmitz, M. Sokolich, K. S. Holabird, T. M. Hazard, M. F. Gyure, and A. T. Hunter }} 
  \bibinfo{journal}{Appl. Phys. Lett.} \textbf{\bibinfo{volume}{99}},
  \bibinfo{pages}{063109} (\bibinfo{year}{2011}).
 
  
\end{thebibliography}
\end{document}